\documentclass[12pt,a4paper]{article}
\usepackage{graphicx}
\usepackage{amsthm}   
 \usepackage[bottom]{footmisc}
\usepackage{amsmath} 
\usepackage{amssymb}  
\usepackage{mathrsfs} 
\usepackage{stmaryrd} 
\usepackage{txfonts} 
\usepackage{xspace}
\usepackage[left=1in,right=1in,top=1in,bottom=1in]{geometry}

\renewcommand\vec[1]{\overrightarrow{#1}}
\newcommand\cev[1]{\overleftarrow{#1}}

\usepackage{enumerate}
\usepackage{appendix}
\usepackage{hyperref} 

\newcommand{\ba}{\begin{eqnarray}}
\newcommand{\ea}{\end{eqnarray}}

\newcommand{\erf}[1]{Eq.~\eqref{#1}\xspace}

\newcommand{\latinit}[1]{\emph{#1}\xspace}

\newcommand{\R}{{\mathbb{R}}}

\newcommand{\C}{{\mathscr{C}}}

\newcommand{\EE}{{\mathscr{E}}}

\usepackage{color}  
\RequirePackage[dvipsnames,usenames]{xcolor}

\newtheorem{example}{Example}

\date{\today}                      

\begin{document}

\title{Optimal high-level descriptions of dynamical systems}

\author{David H. Wolpert\thanks{Santa Fe Institute, 1399 Hyde Park Road, Santa Fe, NM, 87501}\\
\texttt{dhw@santafe.edu}
\and
Joshua A. Grochow\textsuperscript{*} \\
\texttt{jgrochow@santafe.edu}
\and
Eric Libby\textsuperscript{*} \\
\texttt{elibby@santafe.edu} 
\and
Simon DeDeo\textsuperscript{*}\thanks{School of Informatics and Computing, Indiana University, 901 E. 10th St., Bloomington, IN 47408} \\
\texttt{sdedeo@indiana.edu}
}

\maketitle

\begin{abstract}
\noindent To analyze high-dimensional systems, many fields in science and engineering rely on high-level descriptions, sometimes called ``macrostates,'' ``coarse-grainings,'' or ``effective theories''. Examples of such descriptions
include the thermodynamic properties of a large collection of point particles undergoing reversible dynamics, the variables in a macroeconomic model describing the individuals that participate in an economy, and 
the summary state of a cell composed of a large set of biochemical networks. 

Often these high-level descriptions are constructed without considering the ultimate reason for needing them in the first place. Here, we formalize and quantify one such purpose: the need to predict observables of interest concerning the high-dimensional system with as high accuracy as possible, while minimizing the computational cost of doing so. The resulting State Space Compression (SSC) framework provides a guide for how to solve for the {optimal} high-level description of a given dynamical system, rather than constructing it based on human intuition alone. 

In this preliminary report, we introduce SSC, and illustrate it with several information-theoretic quantifications of ``accuracy", all with different implications for the optimal compression. We also discuss some other possible applications of SSC beyond the goal of accurate prediction. These include SSC as a measure of the complexity of a dynamical system, and as a way to quantify information flow between the scales of a system.
\end{abstract}

\clearpage

\setcounter{tocdepth}{2}
\tableofcontents
\clearpage

\section{Introduction}

Historically, scientists have defined the ``level", ``scale", or
``macrostate" of a system in an informal manner, based on
insight and intuition. In evolutionary biology, for example,
a macrostate may be defined by reference to species count, despite a great degree of
within-species diversity, fundamental dependencies between organisms,
and complicated subunits such as tissues and cells which can also
reproduce. In economics, the macrostates of the world's
socio-economic system are often defined in terms of firms, industrial
sectors, or even nation-states, neglecting the internal
structure of these highly complex entities.

How do we know that these choices for the macrostates are the best
ones? Might there be superior alternatives? 
How do we even quantify the quality of a choice of macrostate? 
Given an answer to this question, can we \emph{solve} for the optimal
macroscopic state space with which to analyze a system? This is the general problem of state space compression.

Stated more formally, the first step in State Space Compression (SSC)  is to 
specify a measure of the quality of any proposed map $x_{t} \rightarrow y_{t}$ used to compress a dynamically evolving ``fine-grained" variable
$x_t$ into a dynamically evolving ``coarse-grained" variable
$y_t$.\footnote{Here we will use the term ``coarse-graining"
  interchangeably with ``compression", and so will allow arbitrary
  coarse-graining maps. There is a large literature that instead
  interprets the term ``coarse-graining" to refer to the special case
  where $x$ is the spatial position of a set of $N$ particles and we
  wish to map each $x$ to the spatial position of $m < N$ ``effective
  particles". See~\cite{saunders2013coarse} for a review. There is
  also work that interprets the term ``coarse-graining" to refer to
  the special case where $x$ is a network structure, and we wish to
  map each $x$ to a smaller network
  (see~\cite{itzkovitz2005coarse,blagus2012self}). For discussion of
  the general problem of coarse-graining for computational, cognitive,
  biological and social systems see,
  \emph{e.g.},~\cite{krakauer2010intelligent,dedeo2011effective,dedeo2014group,krakauer2014information,dedeo_thisvol}
  and references therein.}  We will take a utilitarian approach in
deciding what makes for a good compression of the dynamics of a
system: state space compression has a practical benefit, and
quantifying that benefit as an objective function drives our analysis.

Given such an objective function and an evolving fine-grained variable $x_t$, we can try
to \emph{solve} for the best map compressing $x_t$ into a higher-scale
macrostate $y_t$. The dynamics of such an optimally chosen compression
of a system can be viewed as defining its emergent properties. Indeed,
we may be able to iterate this process, producing a hierarchy of
scales and associated emergent properties, by compressing the
macrostate $y$ to a yet higher-scale macrostate $y'$.

In this preliminary report we formalize and quantify one objective
function for compressing a high-dimensional state: The need to predict observables of interest concerning 
the high-dimensional system with as high accuracy as possible, while minimizing the computational cost of doing so. 
Many of the observables of interest that are theoretically and experimentally
analyzed in the sciences are considered interesting precisely because
they make good macrostates according to this objective function: we use them because they lower the
resultant computation and measurement costs while still accurately
predicting a given observable of interest. For instance, in physics,
macrostates of a fine-grained system reference thermodynamic
variables, such as temperature or pressure, that characterize relevant
features which can be measured and used for accurate prediction of
many observables of interest --- at very little computation cost.
Our objective function also provides several natural ways to construct
a hierarchy of scales, extending from less-compressed spaces
(higher computation cost but more accurate predictions) to more-compressed spaces (less
computation cost but less accurate predictions).  For example,
we can use the values in the less compressed spaces to give (perhaps additive) corrections
to predictions from the more compressed spaces.

After introducing this SSC framework and relating it to other work in the literature, we
illustrate it by investigating several information-theoretic quantifications of ``accuracy", all with different implications for the optimal compression. We end with a discussion of some other possible applications of SSC, over and above
the motivational one of improving our ability to predict the future. These include the use of SSC as a measure of the ``complexity" of a dynamic system, and the use of it to quantify information flow among the scales of a system.

\section{State space compression}

\subsection{The benefit of state space compression}
\label{sec:st_sp_compression}

As mentioned, we adopt the view that the benefit of a candidate compression $X \rightarrow Y$ is to
reduce the cost of computing the dynamics over $X$. More precisely,
our job as scientists is to find a compression map $\pi: X \rightarrow Y$, 
dynamical laws $\phi$ for evolution over $Y$, and a map $\rho$ from $Y$ to the observable of interest, such that
the future values $y_t \in Y$ accurately predict the associated observable of interest (defined in
terms of $x_t \in X$), while the three maps $\pi, \phi, \rho$ are relatively easy to compute.

Stated prosaically, we wish to simulate the dynamics of $x_t$, using a simulation that optimally
trades off its computation and measurement costs with its accuracy in predicting some observable of interest concerning
future values of $x$.
From the perspective of coding theory, we are using a lossy code to encode a system's fine-grained state as
an element in a (usually) smaller space, evolving that
compressed state, and then using it to predict
the observables of interest concerning the future fine-grained state.
 
It is important to emphasize that in measuring the quality of a candidate $\pi, \phi, \rho$
we are often not interested in predicting a given microstate $x_t$ \latinit{in toto}. 
Instead, there are certain observables associated with $x_t$ that we are interested in reconstructing. 
An example of this is in studies of bird flocks, where we are often not interested in predicting
where each individual bird is at some future time, but rather the location of the center of mass of birds.
(See  Ex.~\ref{ex:flock} below.)

It is also important to emphasize that there are at least two general types of ``accuracy" concerning our
predictions of such observables such that we
might be interested in. A key issue in many compressions is how \emph{closely} the predicted observable of interest matches its actual value. In such cases, it makes sense to use a distance measure over the values of the observable to quantify the accuracy of predictions. 
When we are not provided such a distance measure exogenously, it may make sense to use
an information-theoretic measure, e.g., a variant of mutual information \cite{coth91,mack03}.
There are other cases though where we are interested not in how accurately a prediction based on
a particular $y_t$ matches a truth based on the associated $x_t$, but rather in how accurately the entire
distribution over (predictions based on) $y_t$ matches the associated distribution based on $x_t$. 
(For example, this is the case when we are interested in accurately predicting future expectation values, or managing measures of risk such as expected variance.)
In these situations a measure like Kullback-Leibler (KL) divergence may be most appropriate.

In this paper will be concerned with the prediction of the entire
future evolution of a system, rather than a value at a single, particular moment in the future. 
This reflects the fact that typically we are interested
in future \emph{behavior}, which is an inherently multi-timestep phenomenon. By requiring that the compression / decompression accurately reconstructs
the time-series as a whole, we ensure that it captures the dynamics of the system.

In addition to SSC's role in the sciences, where it is often implicitly
a central issue in how we choose to analyze a nature-given system,
in engineering the problem of optimal design of a system will very often
reflect the tradeoffs considered in SSC. So results concerning SSC might be helpful in such design --- 
or benefit from what has been learned about such design. 
An obvious example of this central role of SSC's tradeoffs in engineering is in the design of
predictive coding algorithms, e.g., for compressing movies so that they can be streamed 
across the internet to an end-user with minimal bandwidth requirements while still maintaining
high fidelity (as perceived by that end-user).

Another example occurs in CPU design and the problem of branch prediction. In modern CPUs, whenever a conditional ``if-then-else'' statement is encountered, the CPU first guesses---based on certain heuristics and a record of previous branches---which branch of the conditional to take. It starts speculatively computing as though that branch were taken; if it later discovers that it guessed wrong, it terminates the speculative computation and continues along the correct branch, having incurred a delay for guessing incorrectly.

We see here the tradeoff between accuracy and computation cost: it only makes sense to do branch prediction if there is a computationally efficient predictor that is correct most of the time. In any other case, branch prediction would end up wasting more CPU cycles than it saves. Branch prediction is enormously successful in practice, partly because, based on the prior distribution of how most people (or compilers) write and use conditional statements, relatively simple heuristics give (almost shockingly) accurate results. Interestingly, some of the best branch predictors are in fact based on (very) simple models of neurons, as suggested in the discussion below (see, e.g., Ref.~\cite{jimenez2005BP}).

In many cases, maps $x \rightarrow y$ that are good compressions of a dynamical system
in the sense just outlined have the property that the dynamics of $y_t$ is (first-order) Markovian. 
(Indeed, as mentioned below, obeying \emph{exactly} Markovian dynamics is a core component of the definition of a ``valid compression"
considered in~\cite{shalizi2003macrostate,gornerup2010method,gornerup2008method,jacobi2007dual}.)
Even if the best compression is not Markovian,
so long as we are considering a good compression of $x_{t}$ into $y_{t}$, 
once we set the initial value of $y$ often
we can well-approximate the future values of $y_{t}$ with a Markovian process, with 
little further information needed from later $x_{t}$ to correct that dynamics. 
(As discussed below, this is the desideratum defining ``good compression" considered 
in~\cite{israeli2006coarse,israeli2004computational,pfantecomparison}.) For these reasons,
we often restrict attention to compressed spaces whose associated dynamics are Markovian.
This is not a requirement of SSC however.

Finally, we note that compressing the state space of a system to reduce
computation cost while maintaining predictive accuracy
is not only a core requirement of how human scientists and engineers build their models of physical
systems. It is also often a core requirement of those systems themselves. In particular,
it is a guiding principle for how the human nervous system operates. Computation is
very costly for a biological brain, in terms of heat generated that needs to 
be dissipated, calories consumed, etc. Moreover, at an abstract level, the 
(fitness-function mediated) 
``goal" of one's senses and the associated computational processes in the brain is to produce a compressed description of the environment that the brain then uses to produce accurate predictions of the future state of certain fitness-relevant attributes of the environment - all at minimal caloric cost~\cite{clark2013whatever,krakauer2011darwinian,shadmehr2010error}.\footnote{Recent
work has suggested that biological individuality itself---cells, organisms, species---may be understood as the
emergence of new coarse-grained partitions over the biochemical world~\cite{krakauer2014information}.
} So in ``designing" the 
human brain, natural selection is faced with the precise problem of optimal state space compression as 
we have formulated it. This suggests that we may be able to construct powerful 
heuristics for optimal SSC by considering how the human nervous system processes information.

\subsection{Basic concepts and notation}
\label{sec:notation}

Throughout this paper we adopt the convention that upper case variables indicate either a space or 
a random variable defined over that space, with the context making the meaning clear; 
lower case variables indicate elements of a space or values taken by a random
variable, again with the context making the meaning clear. We also adopt the convention of letting the context make clear whether we are talking about a probability distribution or a probability density function, the associated meanings of conditional distributions, and other standard conventions from probability and statistics.

In addition, the measures of the integrals we write will all be implicitly matched to the associated space. 
So when we use the integral symbol for a countable space, we are
implicitly using the counting measure, turning that integral into a sum.

Formalizing the issues discussed in Sec.~\ref{sec:st_sp_compression} leads to the following definitions. The first four are specified exogenously by the problem being studied:
\begin{enumerate}
\item The \textbf{microstate} $x \in X$, which evolves in time according to a stochastic process $P(x_t, \dotsc)$, which in particular specifies a prior distribution over the value of $x$ at a current time $t_0$;

\item A space of possible values of an \textbf{observable} of interest $\Omega$, and an associated 
  observation conditional distribution $\mathscr{O}(\omega \in \Omega \mid x)$, which may be deterministic.\footnote{In addition to allowing stochastic dynamics, whenever it was natural to use a function $f\colon A \to B$ between two spaces, we instead use the more general setting of a conditional distribution $d_f(b | a)$.}

\item\label{item:acc} A (typically) real-valued \textbf{accuracy function} $C : \Omega \times \Omega \rightarrow \R$---usually a metric---quantifies how good a predicted value is compared to the true observable;

\item A \textbf{weight function} $W(t)$ used to determine the relative importance of 
	predictions for the observable at all moments in the future.
\end{enumerate}
When dealing with stochastic processes one may have a distribution over predictions as opposed to a single prediction at each time step. In this case, it may make sense to replace (\ref{item:acc}) with a function that gives some sort of measure of the difference between two probability distributions over $\Omega$, rather than the difference between two points from $\Omega$. 

The remaining three objects are chosen by the scientist, possibly, as we suggest here, by SSC optimization:
\begin{enumerate}
\setcounter{enumi}{4}
\item A \textbf{macrospace} or \textbf{compressed space} $Y$, with elements $y$, called \textbf{macrostates}, that evolve according to a stochastic process $\phi_{t, t_0} (y_{t} \mid y_{t_0})$;

\item A \textbf{compression} distribution, $\pi(y \mid x)$, which compresses $x$ to $y$ (and may be a deterministic function);

\item A \textbf{prediction} distribution $\rho(\omega \in \Omega \mid y)$ which gives the prediction for
the observable based on the compressed state $y$. 
\end{enumerate}
Although there may be many applications in which the dynamics of the microspace and/or the macrospace are first-order Markovian, we do not require this. Furthermore, we also do not require $\phi$ and $P$ to be time-homogeneous. 

We will sometimes refer to the compression distribution as an ``encoding'' distribution, and refer to the prediction 
distribution as a ``decoding'' distribution. We will typically apply the metric $C$ on the observable space to compare the distribution $\rho(\omega \mid y_t)$ with the distribution $\mathscr{O}(\omega \mid x_t)$.

As an example of a microstate process, there could be
a time-homogenous Markov chain with generator $p$, set by Nature, and a marginal distribution at time $0$, $P(x_0)$, set by a scientist (albeit implicitly). The scientist may want to predict the future of the system, the \emph{a priori} probability, with initial condition $x_0$. In this case the full Markovian process is given by 
\ba
P(x_0, x_1, \ldots) &=& P(x_0) \Pi_{t = 1}^\infty p(x_{t} \mid x_{t-1})
\ea



Three terms will contribute to our overall state compression objective function.
The first is the average future value of the accuracy function, 
evaluated under $W(t)$. We call this the \textbf{accuracy cost}. The second term is the
average future \textbf{computation cost} of iterating both $\phi$ and $\rho$. The third term is the cost of evaluating $\pi$ once---or otherwise performing a measurement, e.\,g., of the physical world---to initialize $y$. Although for simplicity we include this in the computation cost, in general it is just a quantification of mapping an initial $x$ to an initial $y$. This may have nothing to do with ``computation", \latinit{per se}. For example, it may reflect a process of physically measuring an initial value $x$, with a noisy observation apparatus
producing a value $z \in Z$ according to a distribution $P(z \mid x)$, where the observed initial value $z$ is then mapped to the starting value of our simulation via a distribution $P(y_0 \mid z)$ that we set. In this case $\pi(y \mid x) = \int dz \; P(y_0 \mid z) P(z \mid x)$, where $P(z \mid x)$ might be more accurately described as a ``measurement cost". Whenever we write ``computation cost,'' it should be understood in this more general manner.

Unless specified otherwise, we take the weight function $W(t)$ to be a probability distribution over $t$.
A particularly important type of $W(t)$ is a simple future geometric discounting function, $W(t) \propto [1- \gamma]^t$.
This is often appropriate when there is \emph{a priori} uncertainty on how far into the future the 
scientist will end up running the computation.{\footnote{Also
see the literature on decision theory, where geometric discounting with a constant $\gamma$ is justified
as necessary to maintain consistency across time in how a decision-maker makes a decision, and especially the literature
on reinforcement learning and dynamic programming~\cite{suba98}.}} As alternatives, we could consider
the (undiscounted) average accuracy cost and average computation cost over some temporal window $t \in [0, T]$.
Another alternative, for computations known to halt, is to consider the average of those two costs from $t = 0$
to the (dynamically determined) halting time. Formally, this means extending the definition of $W$ to be
a function of both $t$ and $y_t$.

We, the scientists, are provided the original stochastic process $P(x_{t'} \mid x_{t})$ and determine the observable $\mathscr{O}$ and weight function $W(t)$ that capture aspects of the system that interest us. Our task is to choose the compression function $\pi$, associated space $Y$, compressed state dynamics $\phi$, and prediction map $\rho$ to the space of observables, such that $\pi$, $\phi$, and $\rho$ are relatively simple to calculate, compared to the dynamics of $x_{t}$ directly, and the resultant 
accuracy costs are minimal (e.\,g., the resultant predictions for $\mathscr{O}(x_{t'})$ for $t' > t$
are minimal distortions of the true values). The best such tuple $\{\pi, Y, \phi, \rho\}$ is the one that 
best trades off the (average) quality of the reconstructed time series with the costs of implementing $\pi$ and calculating
the dynamics of $y_{t}$.
In essence, a good state compression is like a good encoding of a video movie: it produces a compact
file that is simple to decode, and from which we can recover the characteristics of the original movie that 
the human user is attuned to, with high accuracy. 

SSC is a general framework, and its use requires that we make explicit choices for things such as the accuracy and the computation costs. Rather than provide discipline-independent prescriptions, we emphasize that the problem at hand should determine the choices adopted for the various terms described.  For example, one would expect that quite often using several different accuracy
costs would all provide physical insight into what is truly driving the dynamics across $X$.
So in practice, it may be beneficial to consider several accuracy costs, not just one.

\section{Illustrations of our notation} \label{sec:illustrations}

In this section we present several high-level illustrations to help clarify the meaning of our notation and to give an indication of how SSC could be applied in various settings.

\begin{example}

Consider a flock of $N$ labelled birds exhibiting coordinated 
flight~\cite{bialek2012statistical,attanasi2014information,couzin2009collective}. The mass of each bird is fixed. The microstate of the flock, $x$, is the
phase space position of the flock, i.e., the positions and velocities of all $N$ birds; in particular, the space of microstates is then $X = (\R^3)^{2N}$. 
The stochastic dynamics $P(x_t \mid x_0)$ of the microstate
is given by some bird-level rule governing how each bird's acceleration is determined by its current position and velocity as well as
the positions and velocities of other birds in the flock. For the purposes of this example, we assume that 
higher-order derivatives of the positions of the birds---e.g., their accelerations---are not relevant if we
know the positions and velocities of all the birds. Suppose we are interested in predicting the center of mass of the
flock at all future times; thus our observable space $\Omega$ will be $\R^3$, corresponding to the center of mass, and the observation conditional distribution $\mathscr{O}(\omega \mid x)$ is a deterministic function $\mathscr{O} \colon \R^{6N} \to \R^3$ giving the center of mass. We take as our accuracy cost $C\colon \Omega \times \Omega \to \R$ the Euclidean distance between two points in $\Omega = \R^3$.

One way to successfully predict the position of the center of mass is to evolve the stochastic dynamics of the microstate. This
may be computationally challenging for large $N$ since $X = \R^{6N}$.
As an alternative, a successful state space compression of the microstate dynamics would 
be a map from the high-dimensional vector $x$ to some other much smaller-dimensional 
vector of ``sufficient statistics", $y$, such that we can easily and accurately compute the evolution of $y$ into the future and at each future moment
$t$ recover the center of mass of the flock from $y_{t}$. For example, it may be that due to the 
details of the microstate dynamics, a macrostate $y$ comprising 
the $\R^3 \times \R^3 \times (\R^3)^3 = \R^{15}$ vector

$ $

\{Position of the center of mass of the flock; Momentum of the entire flock;

Components of a Gaussian ellipsoid fit to the shape of the flock\},

$ $

\noindent can be easily and accurately evolved forward in time, without concern for any other information in $x$ 
not reflected in those fifteen real numbers.
Since one component of this particular $y$ is precisely the quantity
we are interested in (i.e. the position of the flock's center of mass), we can recover this information from $y_{t}$ with $\rho\colon Y \to \Omega$ which in this case is a projection. Since (by hypothesis) the evolution of $y$ is accurate, this is a good 
state compression, with a small combined value of the computation cost of the dynamics of $Y$ and the Euclidean error of the prediction of the center of mass.
\label{ex:flock}
\end{example}

\begin{example} \label{ex:2}
Consider an agent-based simulation of the behavior of $N$ humans in a set of interacting firms in an
economy evolving in real time, where each player has time-varying observational data concerning the history of
the entire economy and a (time-varying) set of endowments. The microstate $x$ at each moment is the set of associated
observation vectors and endowment sets of each of the $N$ players. 
Let the dynamics of the system be a set of stochastic rules for each of
the players describing what move it makes at each moment as a function of
its current observation vector and endowment set.

Say that what we are interested in is the total GDP of the economy specified in $x$, and
in particular with how that GDP depends on some exogenous parameter of the system that
an external regulator can set. We take the space of observables $\Omega$ to be $\R_{\geq 0}$, representing the total GDP, and the observation map $\mathscr{O}(x)$ to be the total GDP of the economy specified by $x$. We take the accuracy function $C\colon \Omega \times \Omega \to \R$ to be the absolute value of the difference between the simulated and actual GDP.
Even if we posit a precise microscopic stochastic dynamics $x_{t}$, and therefore
 how GDP depends on the exogenous parameter, it
may be very challenging to calculate the GDP for any particular value of the
exogenous parameter..

On the other hand, depending on the details of the microscopic rules for how the
agents behave, it may be that we can coarse-grain $X$ into a compressed space $Y$ that
specifies only the current state of each of the firms---not any individuals within the firms, or non-firm individuals---and that we can both evolve $y \in Y$
accurately and then infer the GDP at each time $t$ very accurately knowing only $y_{t}$. 
In this case the coarse-graining of the economy into characteristics of the firms
in the economy is a good compression of the state space of the economy.

However it may be that this state compression does not work well, giving large expected
error for the prediction of future GDP. It may be that some other state compression, e.g., 
that couples the states of players from multiple firms at once, results in better predictions.
In this example, if we could find that better compression, it would provide major
insight into what drives the evolution of the economy. Working with often highly restricted classes of $\pi$'s and
$\phi$'s this is the aggregation problem of economics~\cite{chiappori2011new,hildenbrand2007aggregation}.

\end{example}

\begin{example} \label{ex:cylinder}
Consider a cylinder of fixed dimensions with a moveable partition separating it into
two half-partitions, with an ideal gas on each side of the partition. Let $z$ be the position
of the partition. Our microspace $X$ will thus be the space of configurations of the particles in each of the ideal gases, together with the position $z$ of the partition. Assume that at some time $t_0$ the partition has a pre-fixed value of $z$,
and there is a pre-fixed temperature and pressure of the two ideal gases, with higher values of
both in the ideal gas lying above the partition, at higher values of $z$. Assume that at time $t_1$ the partition
starts sliding towards smaller values of $z$ at a very slow (essentially adiabatic) constant and pre-fixed rate, stopping
when the system reaches equilibrium. Suppose, however, that we do not know $t_1$; it
is instead distributed according to a Gaussian distribution centered about some particular time.
Let the observable of interest be the temperatures and pressures of the two ideal gases; the observable space $\Omega$ is thus $\R_{\geq 0}^4$ (with temperatures in Kelvin). We take our accuracy function $C\colon \R_{\geq 0}^4 \times \R_{\geq 0}^4 \to \R$ to be Euclidean distance, for simplicity.

Consider the compressed space $Y$ which consists only of the position $z$ of the partition. In this case, given the current value $y_{t}$, we can perfectly predict the future values $y_{t'}$ with $t' > t$, i.\,e., the future positions of the partition,
since we know the rate of movement of the partition. In turn, because we know the dimensions of the cylinder
and the initial temperatures and pressures of the two ideal gases,
we can calculate the temperatures and pressures of the two gases at time $t$ using knowledge of the position of the partition. That is, the prediction map $\rho\colon Y \to \Omega$ computes the pressures and temperatures of the two ideal gases based solely on the position of the partition. (The ``knowledge'' of the initial values of the temperatures and pressures are built in to the map $\rho$, but $\rho$ itself is just a function of $z$, since that is all that is present in the compressed space $Y$.)
So in this case, the compression map $\pi: x \rightarrow y = z$ is perfect, in that it allows perfect prediction of  the future states of the observables that we are interested in.
\end{example}

\section{Related work} \label{sec:related_work}

\subsection{Causal states \& computational mechanics}

While the basic idea of coarse-graining has a long history in the physical sciences, so has the
recognition of the troublingly \emph{ad hoc} nature of many coarse-grainings used
in practice~\cite{Crut87a}. Ref.~\cite{Crut88a} was an important early analysis of
this issue, introducing the notion of causal states, with the associated mathematics known as
computational mechanics~\cite{Crut92c}. 

Causal states have been the workhorse of
a extensive body of of work over the last thirty years; for a recent review, see Ref.~\cite{Crut12a}.
To illustrate the central idea, suppose we are given an ergodic and stationary stochastic process generating
bi-infinite time-series' of the form $\{\ldots, s_{-1}, s_0, s_1, \ldots\}$.
An associated \textbf{causal state} at time $0$ is a maximal set $\epsilon_t$ 
of ``past" semi-infinite strings $s^0_{\leftarrow} \equiv \{\ldots , s_{-1}, s_0\}$ 
such that the probability distribution of any particular
``future" semi-infinite string, $s^0_{\rightarrow} \equiv \{s_1, s_2, \ldots \}$ conditioned on 
any member of $\epsilon_t$ is the same. So all past semi-infinite strings that are in the same causal
set result in the same distribution over possible future semi-infinite strings. 
In the sense that it is maximal, a causal state is an
optimal compression of the original time-series when our goal is perfect knowledge of the future.
The dynamics of the causal states is given by a (stationary, ergodic) unifilar Hidden Markov Model, 
which is called an ``$\epsilon$-machine".
Any $\epsilon$-machine fixes an associated stationary distribution over its causal states. The \textbf{statistical complexity} of 
the machine is  defined as the entropy of that  distribution over the causal states. 

There are several ways one might try to interpret computational mechanics in terms of SSC. 
Neither the dynamics of the variable $s_t$, nor that of the causal states, can be
exactly identified with the dynamics of SSC's fine-grained space.
However because causal states are optimal predictors of the future, the only information they discard is that
which is irrelevant to future prediction. This would suggest that we identify the causal states 
with a coarse-graining close to, but not necessarily identical with, SSC's fine-grained space. 

An alternative interpretation is to identify each past semi-infinite string 
$s^t_{\leftarrow}$ as a fine-grained state $x_t$ of SSC. In particular, since
any such string fixes the associated causal state $\epsilon(s^0_{\leftarrow} )$, it fixes the 
distribution over the sequence of future causal states, and therefore the distribution over all of
the subsequent past strings, $s^{t' > t}_{\leftarrow}$. 
Under this interpretation of computational mechanics, the next value $s_1$ can be viewed as $\omega_0$, a (noisy)
observation of $x_0$, and the causal state $\epsilon(s^0_{\leftarrow} )$ can be seen as $y_0$, a compression 
of $x_0$. Note though that this interpretation of an $\epsilon$-machine always results in an SSC scenario where $X$
is infinite, the dynamics over $x_t$ is ergodic and stationary, and
every $x_{t+1}$ is uniquely determined by $(x_t, \omega_t)$. It does not provide
a way to relate other kinds of SSC scenario to computational mechanics.

Refs.~\cite{shalizi2003macrostate,shalizi2001causal} 
analyze ergodic and stationary bi-infinite strings in terms of $\epsilon$-machines, treating statistical complexity as
a term in an objective function, $V$. If we adopt the second interpretation of computational mechanics in 
terms of SSC suggested above, then the role of statistical complexity in $V$ 
is loosely analogous to the role of the compression term involving $\pi$ in the computation cost of SSC's objective function.
Note though that the term in SSC's objective function based on the computation cost of iterating $\phi$ has no obvious analogue in $V$.
In addition, the analysis in~\cite{shalizi2003macrostate,shalizi2001causal} does not
consider exogenously specified accuracy costs / observation operators.


Nonetheless, the similarity between the work in~\cite{shalizi2003macrostate,shalizi2001causal} 
and SSC suggests that we can extend  computational mechanics
by coarse-graining  causal states. Such  a coarse-graining would mean that
we no longer have maximal predictive power. However it may
result in a more than compensating reduction in computation cost.
Several papers~\cite{Shal99a,creutzig2009past,still2010optimal} have suggested
using the information bottleneck method~\cite{tishby2000information} to coarse-grain
causal states this way, to reduce the complexity of
an $\epsilon$-machine representation. Ref.~\cite{still2014information} provided a thermodynamic
account of this process, and Ref.~\cite{marzen2014circumventing} provides the first explicit treatment for 
$\epsilon$-machines based on time-series data; this latter reference found, among other things, that it was generally more
accurate to derive these coarse-grained machines from more fine-grained models, rather than from the data directly.

This new work quantifies accuracy as the average KL divergence between the optimal
prediction of the future distribution, and that made using a coarse-graining over causal state. (See
the discussion at the end of Sec.~\ref{alternative-it-costs} involving KL divergence as an accuracy cost.)
The goal is to optimize that accuracy given a 
constraint on ``coding cost" (measured by the mutual information between the new, coarse-grained causal 
states and the future distribution). The objective function then becomes (Ref.~\cite{creutzig2009past,still2010optimal}) a linear combination
of coding and accuracy costs, which using the notation of that paper can be written as a minimization of
\begin{equation}
\mathcal{L}(\beta) = I[\cev{X};\mathcal{R}] -\beta I[\mathcal{R};\vec{X}],
\label{minl}
\end{equation}
where $\beta$ provides a single parameter to govern the trade-off between coding costs. 
By optimizing this objective function we find the ``internal states", $\mathcal{R}$ (similar
to the compressed states $Y$ of SSC), that are as
predictive as possible of the future, in that $I[\mathcal{R};\vec{X}]$ is large, while 
at the same time minimizing the coding cost.

The ``memory cost", loosely analogous to SSC's computation cost, is the first term of $\mathcal{L}(\beta)$, $I[\cev{X};\mathcal{R}]$. It measures the coding costs of a soft-clustering of the original causal states; \emph{i.e.}, a probabilistic
map between the original causal states and the coarse-grained space $\mathcal{R}$.  One can also imagine a
hard clustering (\emph{i.e.}, a deterministic many-to-one map) from the original space; in this case, the coding cost reduces to $H(\mathcal{R})$, the statistical complexity of the new
machine.

The second term of $\mathcal{L}(\beta)$ is  analogous to 
the accuracy cost of SSC. In Eq.~\ref{minl}, we maximize mutual information between the model state and the
semi-infinite future; considering all future times of equal prior importance is common choice~\cite{bialek2001predictability,still2014information,wiesner2012information}. 
The more general approach is to consider a weight, $W(t)$, that would make, for example, the near-term future more
important to predict than times that are more distant. In contrast, as described below, in SSC's use of mutual information as an accuracy cost, we iteratively evolve the coarse-grained state further and further into the future, and at each iteration evaluate the mutual information between the coarse-grained state at that moment and the fine-grained state at that moment. 

\subsection{State aggregation, decomposition, and projection methods} \label{sec:related_work_heuristics}

As early as 1961~\cite{simon1961aggregation}, one finds discussions of the tradeoff between accuracy cost, the computation cost of a compressed model, and the cost of finding a good compressed model-- precisely the three costs we consider.\footnote{Where we include in the ``computation cost of a compressed model'' the initial cost of compressing a fine-grained state into a compresed state.}
This led to studies of aggregating or ``lumping'' the states of Markov chains in order to construct compressed (in our language) Markov chains~\cite{auger2008aggregation, simon1961aggregation, chiappori2011new, hildenbrand2007aggregation, white2000lumpable}. Since some systems do not admit any good aggregations, the limitations of aggregation and lumpability methods have recently been fleshed out~\cite{kotsalisShamma1, kotsalisShamma2}.

The ``finite state projection method" (FSP)~\cite{munsky2008finite,munsky2006finite} is a similar method, developed to map a microstate of a chemical
reaction network evolving according to a chemical master equation to a compressed version
to speed up simulation of the evolution of that network. Though designed for evolving such reaction networks, the 
FSP is applicable whenever the set of possible microstates $X$
is countable and the stochastic process generating $x_{t}$ is a Poisson process. 
The idea behind the FSP is to select a large subset $X' \subset X$ and group all the states in $X'$ into one
new, absorbing state. The stochastic process rate constants connecting the elements of $X \setminus X'$
and governing the probability flow from $X \setminus X'$ into $X'$ are not changed. The goal is to choose
$X'$ to be large and at the same time to have the total rate of probability flowing from $X \setminus X'$ into $X'$
be small. While somewhat ad hoc, the FSP has been found to work well in the domain for which it
was constructed.

These same three costs are also important in control theory and reduced-order modeling~\cite{moore1981PCA, lorenz, mori, zwanzig, holmesLumleyBerkooz, chorinHaldBook,antoulas, schilders2008model, beckLallLiangWest, antoulasSorenson, dengThesis, bai2002krylov}. However typically in this work and the aforementioned working on lumpability and FSP the computation cost of the compressed (``reduced") model is  not included in the objective function being optimized, and when it is considered it is typically replaced with the simple proxy of the dimension of the compressed space. 
It is probably true that in general this dimension is not a bad proxy for computation cost.
However there are many specific cases where the actual cost of running the computation in the reduced space would differ considerably from what one would expect based on the dimension of that space. For example, much of this work in control theory
restricts itself to computations in the compressed space that work by (local) linear transformation. 
One would expect that the dimension of the compressed space is often a poor proxy
to the computation cost for such computation,
for certain sparsities, ranks, and condition numbers of the reduced system --- actual computation cost might be smaller with a slightly higher-dimensional compression whose dynamics is significantly sparser, compared to a lower-dimensional but dense one. There are many other properties that might affect the computation cost of evolving the compressed system, both in theory and in practice.  As far as we are aware, these issues have not been considered in any great depth.

While focusing on accuracy cost as aggregation methods do, reduced-order modeling has typically considered algorithms of the following sort (see the above-referenced books and surveys for details). First we compute a subspace of the fine-grained space that we believe captures key features of the dynamics. Typically this is done by performing a variant of the Singular Value Decomposition (SVD - which often goes by the alternative names of ``principal component analysis", ``proper orthogonal decomposition", or ``Karhunen--Loeve expansion") and then choosing the most significant singular directions (corresponding to the largest singular values in the SVD). There is some art in choosing how many directions to project onto, but often a sharp drop-off is observed in the singular value spectrum, and it is this cutoff that is chosen. This somewhat ad hoc, though well-motivated, choice of cutoff is essentially the only place that there is even an
implicit concern for the computation cost of the evolving the reduced system. (Balanced truncation~\cite{moore1981PCA, kotsalisRantzer, kotsalisMegretskiDahleh, balakrishnanSuKoh, lallBeck2003balanced} offers a more principled way to select the cutoff, or ``truncation,'' point, but still does not explicitly take into account the computation cost of the compressed model.) However we choose the subspace, the
next step is to project the dynamics over the fine-grained space onto this subspace (``Galerkin projection'') and then run the associated
(linear) dynamics.

Other methods have also been proposed, under the umbrella term of Krylov iteration, that can frequently make it easier to solve for the optimal compression (again, in terms of accuracy cost), using iterative methods rather than computing the SVD directly. But the end goal, and the way in which accuracy and computation cost of the reduced model are taken into account, are essentially the same in the pure SVD-based methods.

More recently, so-called local SVD (or POD) methods have been proposed~\cite{peherstorferLocal}, in which a model is built as above, but instead of being a model of the whole space, it is only designed to model a local portion of the state space. Many local models are then ``glued'' together, for example, by training a classifier (using standard techniques from machine learning) to decide which of the local models should be used at any given point or trajectory. 

Yet another set of approaches is the Mori--Zwanzig family of methods \cite{mori, zwanzig}. These are most
naturally viewed as an expansion of the (fine-grained) dynamics in terms of time, truncating the expansion beyond a certain number of steps in the past. 
However the survey of Beck, Lall, Liang, and West \cite{beckLallLiangWest} emphasizes the close relationship between Mori--Zwanzig and the methods of reduced-order modeling discussed above. Polynomial chaos, used in uncertainty quantification, also has a similar flavor: at a high level, it involves expanding in terms of certain nice polynomials and then truncating this expansion.\footnote{Technically, polynomial chaos does a compression of a probability distribution on the fine-grained space, rather than compressing the fine-grained space itself. But this compression of the probability distribution is still implemented in terms of truncating an expansion.}

Israeli and Goldenfeld \cite{israeli2006coarse} also consider the relationship between coarse-graining and the computational cost of prediction, in the context of cellular automata. They consider certain spatial coarse-grainings, which are required to ``commute" 
with the micro-dynamics, in
that if one evolves the microdynamics and then coarse-grains, this gives the same result as coarse-graining
first and then and evolving the coarse dynamics. They show that even this highly restricting class of compressions of cellular automata can make certain ``computationally irreducible'' automata easier to predict. They generally measure ease of prediction by a qualitative measure: for example, certainly the CA became easier to predict if it was initially Turing-complete, but the compressed version is not. It would also be natural in this setting to provide a more quantitative measure of ``ease of prediction.''
However, in addition to only considering a limited class of coarse-grainings, they also do not consider an exogenously specified accuracy cost. Rather, they consider any valid coarse-graining of a CA to be giving some information about (the dynamics of) that CA.
Despite these qualifications, this is nonetheless a particularly interesting result suggesting that even limited kinds of state-space compression can be successful in reducing the computational cost of prediction.

We suspect that all of these methods will prove useful in certain instances of state-space compression.
However it is not immediately obvious how to incorporate more general (or more nuanced) notions of computation cost directly into these methods.

\subsection{Generally applicable accuracy costs}
\label{sec:sort_of_related_work}

Another thread of work has tried to define whether a map $x_{t} \rightarrow y_{t}$ is a ``valid" compression, 
without trying to rank such maps or find an optimal solution. For example, the work in
~\cite{gornerup2010method,gornerup2008method,jacobi2007dual,derisavi2003optimal,derisavi2007symbolic} 
starts with the set of four variables $x_0$, $y_0$, $x_t$ and $y_t$, where $t$ is some fixed value greater than $0$, and $y_0$ is
produced by applying a proposed compression map to $x_0$, while $y_t$ is produced by applying that same map to $x_t$.
It then considers $y_t$ to be a valid compression of $x_t$ 
if the associated dynamics $y_{t}$ is (first-order) Markovian. 

The work in~\cite{israeli2006coarse,israeli2004computational,pfantecomparison}
is also concerned with these four variables, $x_0$, $y_0$, $x_t$ and $y_t$.
Here the stochastic relationship of these four variables is used to assign 
a real-valued quality measure to the map $x_t \rightarrow y_t$, rather than specify whether it is (not)
a valid compression. This quality measure is based on the amount of
extra information that is needed from $x_{0}$, in addition to the value $y_{0}$, for us to accurately predict $y_t$. 
One difference between this work and SSC in that this work does not take computation or measurement cost into account. 

In~\cite{woma00}, optimal compression was implicitly defined in terms of how accurately a compressed
description of a system could predict the fine-grained description. Related work in~\cite{woma07} implicitly
defined optimal state compression in terms of how different probability distributions were at coarse-grained and
fine-grained scales. As with the works discussed above, these works do not consider computation cost,
at least not directly.

These previous studies related to state compression makes compelling points, and generally accords with intuition. However 
one striking feature of this previous work is that none of 
it considers what a state compression of a fine-grained dynamics $x_{t}$ is \emph{for}. 
%
As a result, some aspects of this earlier work makes no sense from an SSC perspective when
applied in certain scenarios.
For example, if $y_{t}$ is just a constant, not evolving in time, then
the dynamics $y_{t}$ is perfectly Markovian, of first-order. So this state space compression, $x \rightarrow constant$, is a ``valid"
compression,  according to some of this  
earlier work. 

Similarly, suppose that extra information from the fine-grained value 
$x_0$ cannot provide extra help in predicting the future
value $y_{t}$ beyond just knowing $y_0$. This would imply that extra information about
$x_t$ cannot provide extra help in predicting the future
value $y_{t}$ beyond just knowing $y_0$. It might seem that in this case $x \rightarrow y$
should be deemed a ``good" state compression.  
After all, if extra future information about future fine-grained states $x_t$
cannot help us predict future compressed states, then dynamics in the compressed space  is ``autonomous", 
essentially independent of $x_t$. 
This is the motivation for much of the analysis in~\cite{pfantecomparison}, for example.

However this reasoning differs in important ways from the motivations of SSC. In particular, say we used this reasoning to argue 
along with~\cite{pfantecomparison} that we have a good state space compression $\pi: x \rightarrow y$ if 
the conditional mutual information $I(y_{t} ; x_0 \mid y_0)$ is small, i.e., if knowing
$x_{0}$ does not help us predict $y_{t}$ any more than knowing $y_0$ does. With this criterion, we would say that
the compression map that sends $x$ to a uniformly noisy value of $y$, which is statistically independent of $x$, is a ``good
state space compression"; it results in $I(y_{t} ; x_0 \mid y_0) = 0$. In contrast, most SSC objective functions
would not assign a high value to such a $\pi$ .

There are also important features of SSC's focus on the full compression / decompression
loop that are absent from this earlier work. For instance, the earlier work considers only the compression $\pi$, with
no associated ``decompression" map $\rho$ that maps $Y$ to an observable of interest. In contrast, we consider the case
where one is interested in ``decompressing" future values of $y$, to make predictions of
observable functions of $x_t$. In addition, this earlier work assumes that future values of $y_t$ are obtained
by iteratively applying $\pi$ to $x_t$. Instead, we allow dynamics in $y_t$ to
evolve according to an arbitrary map $\phi$ from an initial value $y_0 = \pi(x_0)$.
This means that rather than just assign value to a compression map $\pi$,
we assign value to a triple $(\pi, \phi, \rho)$.{\footnote{To state this more formally, 
note that in this earlier work the Bayes net relating the four variables is 
$P(y_t \mid x_t) P(y_0 \mid x_0) P(x_t \mid x_0)$. In contrast,
the Bayes net relating those four variables in SSC is
$P(y_t \mid y_0) P(y_0 \mid x_0) P(x_t \mid x_0)$. This difference reflects the fact that this earlier work
is ultimately concerned with different issues from those central to SSC.}}

\section{Accuracy costs} \label{sec:acc_costs}
In this section, we show how one can quantify accuracy cost (i.\,e., negative reconstruction accuracy).
We do this both when an accuracy function is exogenously provided, and
when it is not --- in which case it may make sense to use an information-theoretic accuracy cost.

\subsection{Exogenously provided accuracy costs}

We begin with an example of accuracy cost appropriate for the case that the dynamics are Markovian. 
\begin{eqnarray}
\EE(\pi, \phi, \rho ; P) &\equiv& \int_{\Delta t > 0} d \Delta t \int dx_0 \; dx \; d\omega \; dy_0 \; dy \; d\omega'  \nonumber \\
&& \;\;\;\; \; W(\Delta t)
 P(x_0) \pi(y_0 \mid x_0) P_{\Delta t}(x \mid x_0) {\mathscr{O}}(\omega \mid x) \phi_{\Delta t}(y \mid
    y_0) \rho(\omega' \mid y)  C(\omega, \omega')  \nonumber \\
\label{eq:curly_C_def}
\end{eqnarray}
In the integrand, $C(\omega, \omega')$ is the cost if our simulation using the compressed space predicts
that the observable has the value $\omega'$ when it is actually $\omega$. $\rho(\omega' \mid y)$
is the distribution (typically a single-valued function) for how we convert the state $y$ of the
compressed space into such a prediction of the value of the observable. $\phi_{\Delta t}$ and $P_{\Delta t}$
are the first-order Markov chains of our simulation over the compressed space, and of the actual 
fine-grained system state, respectively. $\pi$ is the distribution by which we compress the initial state of the fine-grained
state, and finally, $P(x_0)$ is the \emph{a priori} probability that we will be interested in 
simulating the fine-grained system that starts in state $x_0$.

Several variants of Eq.~\eqref{eq:curly_C_def} for accuracy cost are possible, even in the Markovian case. For example, one might be interested in a worst-case accuracy cost over the initial states, rather than an average. More generally, if
the fine-grained dynamics and/or our simulation are not first-order Markovian, then Eq.~\eqref{eq:curly_C_def} would have to be modified accordingly. (We don't present that modification here because in general it can be very messy.)

\subsection{Information-theoretic accuracy costs} 
\label{sec:info_theory_obj}

If an accuracy function $C$ is not supplied and is not obvious to construct, it may be appropriate to replace Eq.\eqref{eq:curly_C_def}
with an information-theoretic definition of accuracy cost. Similarly, if in the problem at hand it's more natural to compare the \emph{distribution} over values of predicted observables 
with the distribution over values of actual observables, then again a (different) information-theoretic definition of accuracy cost may be appropriate. We consider both of those variants of Eq.\eqref{eq:curly_C_def} in the rest
of this subsection. 

\subsubsection{Accuracy cost based on mutual information for two time-steps}
\label{sec:acc_cost_mutual_two_time}

We begin by focusing on the special case in which there are only two time steps, $t_0$ and $\Delta t$. 
Suppose that we know $\omega'_{\Delta t}$ and want to treat that
value as a prediction of $\omega_{\Delta t}$. A natural information-theoretic measure is the mutual information between the predicted value of $\omega_{\Delta t}' \in \Omega$ made after state compression and the actual future value of the observable $\omega_{\Delta t}$, generated by the Markov process $P$ over $X$ and the observable distribution $\mathscr{O}$.

Although intuitively straight-forward, the fully formal equation for this accuracy cost is a bit complicated.
This is because the random variables whose mutual information we
are evaluating are coupled indirectly, through an information channel that goes through the time-0 conditional distribution $\pi$.
Writing it out, this accuracy cost is:
\begin{eqnarray}
\EE_{\Delta t}(\pi, \phi, \rho ; P) &=& -I_{{\mathscr{P}}_{\pi, \phi, \rho ; P}}(\Omega'_{\Delta t}\; ; \Omega_{\Delta t}) 
\label{eq:info_theory_ext}
\end{eqnarray}
where the negative sign reflects the fact that large mutual information corresponds to low misfit $C$,
and where the joint probability ${\mathscr{P}}_{\pi, \phi, \rho ; P}(\omega'_{\Delta t}, \omega_{\Delta t})$ defining 
the mutual information at time $\Delta t$ is given by the marginalization

\begin{eqnarray}
{\mathscr{P}}_{\pi, \phi, \rho ; P}(\omega'_{\Delta t}, \omega_{\Delta t})   &\equiv& \int dx_0 \; P(x_0) 
    {\mathscr{P}}_{\pi, \phi, \rho ; P}(\omega'_{\Delta t}, \omega_{\Delta t}, x_0)
\label{eq:curly_P_def1}
\end{eqnarray}
where ${\mathscr{P}}_{\pi, \phi, \rho ; P}(\omega'_{\Delta t}, \omega_{\Delta t}, x_0)$ is defined as
\begin{eqnarray}
\int dy_0 dy_{\Delta t} \;  
 \pi(y_0 \mid x_0) \phi(y_{\Delta t} \mid y_0)  \rho (\omega'_{\Delta t} \mid y_{\Delta t}) P(x_{\Delta t} \mid x_0) \mathscr{O}(\omega_{\Delta t} \mid x_{\Delta t}). \nonumber \\
\end{eqnarray}
Intuitively, the distribution ${\mathscr{P}}_{\pi, \phi, \rho ; P}(\omega'_{\Delta t}, \omega_{\Delta t}, x_0)$ 
couples $\omega'_{\Delta t}$ and $\omega_{\Delta t}$ by stochastically inferring $y_{\Delta t}$ from $\omega'_{\Delta t}$, then ``backing up'' from $y_{\Delta t}$ to $y_0$ and on to $x_0$, and finally evolving forward stochastically from $x_0$ to get an $x_{\Delta t}$ and thereby $\omega_{\Delta t}$.{\footnote{By the data processing equality, including $\rho$ in this definition cannot increase the associated mutual
information, and arguably it should be removed, as essentially spurious. 
That would result in a mutual information between $Y$ and $\Omega$,
not between $\Omega'$ and $\Omega$. For pedagogical simplicity though, here we look at the mutual
information involving $\Omega'$, the prediction made via $\rho$.}}


When there is no exogenously specified observable, we may simply take the microstate to be the observable, and ask about the mutual information between the future microstate and the future macrostate. This is essentially why mutual information has been used in other work related to SSC~\cite{shalizi2003macrostate,shalizi2001causal,israeli2006coarse,israeli2004computational,pfantecomparison}. However our information-theoretic accuracy cost should be distinguished from these other information-theoretic costs in temrs of \emph{which} mutual information is considered, and what it is conditioned on.

In~\cite{pfantecomparison} the analogous accuracy cost,
defined for the values of a process at a pair of times $t_0$ and $t_1 > t_0$, is the conditional mutual information
$I(Y_{\Delta t} ; X_{t_0} \mid Y_{t_0})$. Although there are scenarios in which both this cost and the cost $\EE_{\Delta t}(\pi, \phi; P)$ in \erf{eq:info_theory_ext} are minimal,\footnote{For example, this occurs if all of the conditional distributions $\pi$, $\phi$ and $P(x_{\Delta t} \mid x_0)$ are deterministic, measure-preserving functions, so that the dynamics in $y$ uniquely specifies dynamics in $x$.} 
there are also scenarios in which the cost $I(Y_{t_1} ; X_{t_0} \mid Y_{t_0})$ achieves its minimal value even
though the cost $\EE_{\Delta t}(\pi, \phi ; P)$ is \emph{maximal}. For example, the latter occurs if
$\pi$ is pure noise, so that dynamics in $y$ implies nothing whatsoever about dynamics of $x$.\footnote{This
distinction between these two measures reflects the fact that they are motivated by different desiderata. The cost $I(Y_{\Delta t} ; X_{t_0} \mid Y_{t_0})$ is motivated by the observation that if it is zero, then
there is no extra information transfer from the dynamics of $X$ that is needed to predict the
dynamics of $Y$, once we know the initial value $y_{t_0}$, and in this sense dynamics in
$Y$ is ``autonomous" from dynamics in $X$.}

\subsubsection{Accuracy cost based on mutual information for more than two time-steps}

The natural extension of Eq.~\eqref{eq:info_theory_ext} to multiple times is
\begin{eqnarray}
\EE(\pi, \phi, \rho ; P) &=& \int d\Delta t \; W(\Delta t) \EE_{\Delta t}(\pi, \phi, \rho ; P) \nonumber \\
& = & -\int d\Delta t \; W(\Delta t) I_{{\mathscr{P}}_{\pi, \phi, \rho ; P}}(\Omega'_{\Delta t}\; ; \Omega_{\Delta t}) 
\label{eq:wrong_mut_info_acc_cost}
\end{eqnarray}
with ${\mathscr{P}}_{\pi, \phi, \rho ; P}(\omega'_{\Delta t}, \omega_{\Delta t})$ defined as in~\erf{eq:curly_P_def1}
for all values of $\Delta t$. 
However, the following example illustrates a subtle but important problem with this formula:

\begin{example} 
Consider
%
a discrete-time system with $X = \{0, 1\}$ with dynamics $P(x_{t+1} \mid x_t) = \delta_{x_{t+1}, x_t}$, and let $Y = X$ but with non-stationary dynamics that swaps the two values at every time step. Suppose $\pi\colon \{0,1\} \to \{0,1\}$ is the identity map. Then at the initial time $t_0$, the map $\rho_{even}\colon Y \to X$ defined by $\rho_{even}(0) = 0$ and $\rho_{even}(1) = 1$ is a perfect predictor of $x_{t_0}$ from $y_{t_0}$; indeed, this same predictor works perfectly at every time that is an even number of steps from $t_0$. At those times $t$ that are an odd number of steps from $t_0$, $x_t$ can still be perfectly predicted from $y_t$, but now by a different map $\rho_{odd}\colon Y \to X$, which swaps the two values ($\rho_{odd}(0) = 1$ and $\rho_{odd}(1) = 0$). In such a situation, mutual information is maximal at all moments in time. However, there is no single, time-invariant map $\rho$ that allows us to interpret $y_t$ as a perfect prediction for the associated $x_t$.

\end{example}

One way to resolve this problem 
is to modify that accuracy cost to force the prediction map from $Y$ to $\Omega$ to be time-invariant. To state this formally, return to the motivation for using information theory in the first place: 
by constructing a space of codewords and an associated (prefix-free) coding function that allow us
to map any value in $Y$ to a value in $\Omega$, taking as our accuracy cost the minimal expected length of those codewords (over all codes). To make this expected codelength precise, we need to define an encoding function. So construct a $Z$ 
and a (time-invariant) encoding function $f$ such that for any $y$, $\omega$, there is a $z \in Z$ such that $f(y, z) = \omega$. From one $t$ to the next, given $y_t$, we have to choose a $z_t$ so that we can recover $\omega_t = \mathscr{O}(x_t)$ by evaluating (the time-invariant) function $f(y_t, z_t)$.
We then want to choose a code for $Z$ that minimizes the length of (codewords for) $z$'s that allow us to recover $x$ from $y$, averaged over time according to $W$ and over pairs $(\omega'_t, \omega_t)$ according to $\mathscr{P}_{\pi,\phi,\rho;P}(\omega'_t, \omega_t)$. 

So we are interested in the $t$-average of expectations of (lengths of codewords specifying) $z$'s where those expectations are
evaluated under $\mathscr{P}_{\pi,\phi,\rho;P}(x'_t, x_t)$. This is just the expectation under the single
distribution given by $t$-averaging the distributions $\mathscr{P}_{\pi,\phi,\rho;P}(x'_t, x_t)$. Write
that single $t$-averaged distribution as 
\ba
{\overline{\mathscr{P}}}_{\pi,\phi,\rho}(\omega', \omega) &\equiv& \int dt \; W(t) \mathscr{P}_{\pi,\phi,\rho;P}(\omega'_t, \omega_t)
\ea
The associated 
minimum of expected codelengths of $z$'s is just $H_{\overline{\mathscr{P}}_{\pi,\phi,\rho}}(\Omega \mid \Omega')$. 
To normalize this we can subtract it from the entropy of the marginal, $H_{\overline{\mathscr{P}}_{\pi,\phi,\rho}}(\Omega)$. (Note that this entropy of the marginal is fixed by $P$, independent of $\pi, \phi$ or $\rho$.) This gives us the change in the expected length of codewords for 
specifying values $\omega_t$ that arises due to our
freedom to have those codewords be generated with a different code for each value of
the prediction $\omega'_t$. Since higher accuracy corresponds to lower accuracy cost, this motivates  
an information-theoretic accuracy cost given by 
\ba
\C(\pi, \phi, \rho; P) &=& -I_{\overline{\mathscr{P}}_{\pi,\phi,\rho}}(\Omega' ; \Omega)
\label{eq:mut_info_acc}
\ea
This information-theoretic accuracy cost is the mutual information under the $t$-average of
$\mathscr{P}_{\pi,\phi,\rho}(\omega'_t, \omega_t)$, rather than~\erf{eq:info_theory_ext}'s $t$-average of the mutual information under the individual $\mathscr{P}_{\pi,\phi,\rho}(\omega'_t, \omega_t)$'s.

\subsubsection{Alternative information-theoretic accuracy costs}
\label{alternative-it-costs}

While it seems that the distribution 
${\mathscr{P}}_{\pi, \phi, \rho ; P}(\omega'_{\Delta t}, \omega_{\Delta t})$ is often a good one
to consider, and often we will want to use an information-theoretic accuracy
cost, in some circumstances we may not be directly concerned with the mutual information
of ${\mathscr{P}}_{\pi, \phi, \rho ; P}(\omega'_{\Delta t}, \omega_{\Delta t})$. 

For example, in accuracy costs that are not information-theoretic, all we are concerned with is the
average discrepancy between the prediction $\rho(y_t)$ and
the truth $\omega_t$. We do not try to ``normalize" that average discrepancy. If it so happens 
that the distribution over $\omega_t$ is close to a delta function, independent of $x_t$, and $\rho(y_t)$ 
just so happens to equal the
center of that delta function, we typically say that the accuracy of the prediction is high;
we do not try to normalize for the fact that $\omega_t$ could be accurately predicted even
without access to the value $y_t$, due to the fact that it is generated by a distribution that is
almost a delta function.

In the context of our accuracy cost based on expected codelengths, this suggests that we
not try to normalize the conditional entropy, $H_{\overline{\mathscr{P}}_{\pi,\phi,\rho}}(\Omega \mid \Omega')$,
by subtracting it from $H_{\overline{\mathscr{P}}_{\pi,\phi,\rho}}(\Omega)$. In other words, to most
closely align with what the term ``accuracy cost'' means in the context of accuracy costs that are not information-
theoretic (i.e., based on accuracy functions),
we may want to use that conditional entropy as our information-theoretic accuracy cost, rather than the 
related mutual information, $-I_{\overline{\mathscr{P}}_{\pi,\phi,\rho}}(\Omega ; \Omega')$. To
be precise, in some circumstances we may want to use
\ba
\C(\pi, \phi, \rho; P) &=& H_{\overline{\mathscr{P}}_{\pi,\phi,\rho}}(\Omega \mid \Omega')
\ea
This accuracy cost will be small if and only if for every prediction $\omega'_t$ 
that arises (with significant probability) there is always a unique associated $\omega_t$ that occurs 
(with high probability, and averaged over times $t$). Whether or not you could far more accurately make
that prediction by using $\omega'$ than if you didn't have access to $\omega'$ is irrelevant --- after
all, you \emph{do} have access to it.

Another example of why we may not want to use~\erf{eq:mut_info_acc} to quantify an accuracy cost
is the well-known fact that 
mutual information in general has a substantial ``artifact" arising via the prior distribution over either of
its two random variable arguments.{\footnote{As an 
extreme example, if we have covarying random variables $A$
and $B$ and $A$ is almost constant, then the mutual information 
$I(A ; B)$ is very close to zero, even if the conditional
distribution $P(b \mid a)$ is highly accurate.}} This is one of the reasons that 
people often replace mutual information with measures like channel capacity 
(which, like $H_{\overline{\mathscr{P}}_{\pi,\phi,\rho}}(\Omega \mid \Omega')$,
only depends on the conditional distribution $\mathscr{P}_{\pi,\phi,\rho}(\omega_t \mid \omega'_t)$).

As a final example of why we may not want to use~\erf{eq:mut_info_acc}, 
recall it was motivated by presuming that ``... we know $\omega'_{\Delta t}$ and want to treat that
value as a prediction of $\omega_{\Delta t}$". This use of compressed space computation motivated
the accuracy costs discussed above. However there are other reasons to
run a computation over a compressed space rather than original space. Some of these
are naturally formulated using information-theoretic accuracy costs. 

An important example of this is when
the microstate Markov process $P_{\Delta t}(x \mid x_0)$ (or the observable $\mathscr{O}$, for that matter) is not deterministic, and
our goal is to use $x_0$
to predict the future evolution of the \emph{entire distribution} over $\omega_{\Delta t}$ given by $\mathscr{O}(\omega_{\Delta t} \mid x_{\Delta t}) P(x_{\Delta t} \mid x_0)$, rather than predict
the specific future values $\omega_{\Delta t}$. In particular, in many 
situations it may prove useful to use Monte
Carlo sampling of the distribution over $Y$ values, ${\mathscr{P}}_{\pi, \phi, \rho ; P}(\omega'_{\Delta t}, x_0)$, as an approximation to Monte Carlo sampling of $\mathscr{O}(\omega_{\Delta_t} \mid x_{\Delta t}) P(x_{\Delta t} \mid x_0)$. (For example, this is often
the case in the kinds of situations where
we might want to use particle filters or some of the techniques of uncertainty quantification.)
A natural accuracy cost for this kind of situation is
\begin{eqnarray}
\C(\pi, \phi, \rho ; P) &\equiv&
-\int_{\Delta t > 0} d \Delta t  \; W(\Delta t) \int dx_0 \; P(x_0) \text{KL}[{\mathscr{P}}_{\pi, \phi, \rho ; P}(\Omega'_{\Delta t}  \mid x_0) \; || \;
{\mathscr{P}}_{\pi, \phi, \rho ; P}(\Omega_{\Delta t}  \mid x_0)] \nonumber \\
\label{eq:KL_distance_acc_cost}
\end{eqnarray}
where the notation ``$\text{KL}[P(A \mid b)  \; || \; R(A)]$" means the Kullback--Leibler divergence between the two distributions
over values $a \in A$ given by $P(a \mid b)$ and $R(a)$~\cite{coth91,mack03}.\footnote{Note that there is
not the same issue here involving dynamic changes to how we match each element of $y$
with an element of $x$ that arose in our analysis of accuracy cost based on mutual information. The reason
is that both of the distributions in the Kullback--Leibler divergence are defined over the exact same space.}

\section{Computation costs} \label{sec:comp_costs}
The core concern of SSC is how to choose $\pi, \phi$, and $\rho$ in a way that
minimizes computation cost without sacrificing too much accuracy cost. To quantify this goal 
we need to quantify the computation cost associated with any tuple $(\pi, \phi, \rho; P)$ (with
the associated $X, Y$ and $\mathscr{O}$ being implied). This section discusses possible quantifications. We emphasize again that we are not advocating for any one particular quantification of computation or measurement cost, nor even that one be selected from among the list we consider here. As with most aspects of the SSC framework, the computation/measurement cost should be selected appropriately for the problem at hand.

We consider three sources of motivation for the computation costs we present: (1) theoretical computer science, (2) information theory, and (3) pragmatism. Although theoretical computer science, and the field of computational complexity in particular, has had a solid quantification of computational resources since the 1960s \cite{hartmanisStearns}, the types of quantifications considered do not lend themselves easily to application of optimization techniques. Thus we also consider costs motivated by information theory that we hope are more tractable for optimization.

As a final comment, we note that even if we wish to use a mathematical expression for computation cost, often
it makes sense to be purely pragmatic, and ignore many of the subtleties discussed above.
This is discussed below, in Sec.~\ref{sec:heuristics}.

\subsection{Defining computation cost in terms of \texorpdfstring{$X$, $Y$, $\pi$, $\phi$, and $\gamma$}{X, Y, pi, phi and gamma}}
\label{sec:interpretation_for_comp_cost}

Suppose we are concerned with an arbitrary computation, independent of any consideration of state compression.
In this case $x  \in X$ would be a specification of some initial data of that computation, $Y$ would be the state of
our computer, $\pi$ would be a function that uses $x$ (in combination with a specification
of the computer program) to initialize the state of the computer, and $\phi$ would be the
dynamics of the computer. A halt state of the computer is a fixed point of $\phi$.


The combination of the computer architecture and the compiler determine both $\pi$ and $\phi$. So
changing either one would change $\pi$ and $\phi$, even if they do not change
the quantity ultimately computed, i.e., do not change the map from an initial
data set $x$ to  an associated attractor in $Y$. In general, all those choices of $\pi$ and $\phi$  that
result in the same ``quantity ultimately computed" will have
different computation costs, as the term is used below. The same quantity can be computed using
many different programs, all differing in the cost to run those programs.

\subsection{Computation cost measures based on theoretical computer science}

There are many ways to quantify computation cost. Indeed,
quantifying scaling relationships among some of the different
kinds of computation cost is a core concern of the entire field of computational 
complexity~\cite{moore2011nature,hopcroft2000jd}.

One of the kinds of computation cost considered in computational complexity is
the running time of the computation. This is also often a primary concern
of real-world SSC, where we are interested in expected ``wall-clock'' time 
of a simulation. If we restrict attention to von Neumann architectures, then for many purposes this cost can be lower bounded by the sum of the expected codelength of messages that a CPU sends to its RAM, over all iterations of the computer. 

As a practical issue, this measure is often accessible via profiling of the program ($P$ or $\phi$ as the case might be)
that is running on the computer. This can then be used to guide the search for a 
$(\pi,\phi,\rho)$ that optimizes the trade-off between computation cost and accuracy cost (see Sec.~\ref{sec:full_monty}).

Often, though, we want a more broadly applicable specification of computation cost, represented
as a mathematical expression; at a minimum this is needed for any kind of mathematical
analysis. One obvious way to do this is to use 
Algorithmic Information Content (AIC), i.e., to quantify computation cost
as the minimal size of a Turing machine that performs the desired computation~\cite{chaitin2004algorithmic}. 
However as has often been remarked, AIC 
has the major practical problem that how one measures the size of a Turing machine $\mathscr{T}$ (i.e., what
universal Turing machine one chooses to use to emulate the running of $\mathscr{T}$) is essentially arbitrary. 
Furthermore, AIC is formally uncomputable, so one has to settle for results concerning asymptotic behavior.
To get around this issue, people sometimes ``approximate'' the AIC of a string, e.g., with its
length after Lempel--Ziv compression. However this in essence reduces AIC to Bayesian maximum \latinit{a posterior}
coding, where the prior probability distribution is implicit in the Lempel--Ziv algorithm. (There are also further problems in using either Lempel--Ziv---see, e.\,g., \cite{shalizi-lempel-ziv-2003}---or AIC---see, e.\,g., \cite{ladyman2013complex}.)

There are several reasonable variants of AIC that might also be appropriate
for some types of analysis of SSC. One set of such variants are the many versions of Rissanen's
(minimum) description length (MDL~\cite{rissanen1983universal,barron1998minimum}). Another one, quite close to the measure of running time mentioned above, is logical depth~\cite{bennett1995logical}. However logical depth  is still subject to the practical difficulties associated with AIC.

\section{The full SSC objective function} \label{sec:full_monty}

Formally speaking, once we have defined both an accuracy cost function and a computation
cost function, we are faced with a multi-objective optimization problem of how to choose $\pi, \phi$,
and $\rho$ in order to minimize both cost functions. There are many ways to formalize this problem.
For example, as is common in multi-objective optimization, we might only wish to find
the set of triples $(\pi, \phi, \rho)$ that lie on the Pareto front of those two functions.
Alternatively, we might face a constrained optimization problem. For example, we might
have constraints on the maximal allowed value of the accuracy cost, with our goal
being to minimize computation cost subject to such a bound. Or conversely we might have constraints on the maximum allowed computation cost (say, in terms of minutes or dollars), with our goal being to minimize accuracy cost subject to such a bound.

In this paper, for simplicity we will concentrate on ways to reduce the multi-objective optimization
problem into a single-objective optimization problem. To do 
this requires that we quantify the trade-off between computation and accuracy costs
in terms of an overall SSC objective function that
we want to minimize. Such an objective function maps any tuple $(\pi, \phi, \rho; P)$ (with
the associated $X, Y$, and $\mathscr{O}$ being implied) into the reals. The associated goal of SSC is to solve for
the $\pi, \phi$ and $\rho$ that minimize that function, for any given $P, X$, and $\mathscr{O}$.

\subsection{The trade-off between accuracy cost and computation cost} 
\label{sec:trade_offs}

Perhaps the most natural overall SSC objective function
is simply a linear combination of the computation cost and  accuracy cost:
\ba \label{eqn:objective}
{K}(\pi, \phi, \rho; P)  &\equiv& \kappa {\C}(\pi, \phi, \rho; P) \;+\; \alpha \EE(\pi, \phi, \rho ; P)
\ea

As mentioned throughout the paper, the computation/measurement cost $\C$ can (in many situations, should) include the cost of mapping the original state to the compressed state. In this case, the computation cost $\C$ might include a term of the form, e.\,g., $H_{\pi, P}(Y_0)$. When all these costs are defined information-theoretically, this quantity has a nice interpretation as the minimum of the expected number of bits that must be transmitted to ``complete the compression-decompression circuit'', i.\,e., the average number of bits
needed to map 
\ba
x_0 \rightarrow y_0 \rightarrow y_t \rightarrow \omega'_t.
\ea

There are some interesting parallels between these objective functions and various ``complexity measures''
that have been proposed in the literature to map a (bit string representation of) infinite time-series $x_t$ to a real number, in particular those that are based on Turing machines.
The cost of computing $\pi$ can be viewed as (analogous to) the length of a Turing machine that takes in $x_0$ and produces $y_0$. The remainder of the computation cost can be viewed as analogous to 
the time it takes to run a Turing machine for the dynamics of $y_0$. Finally, the accuracy cost term can be viewed as analogous to the amount of extra information that must be added to the result of running that Turing
machine to generate (an approximation to) the full observable time-series of interest, $\omega'_t$.

So if we were to just minimize the computation cost of $\pi$, the resultant value is analogous to the algorithmic information content of (an infinite string representation of) the time series of all values $x_t$. Minimizing the combined computation cost of $\pi$, $\phi$, and $\rho$ is instead analogous to the logical depth of $\omega'_t$. On the other hand, the sum of the cost of $\pi$ and the cost of $\rho$ is analogous to the ``description length'' of the time series $\omega'_t$~\cite{rissanen1983universal,barron1998minimum}. 
So minimizing the sum of those two terms is analogous to using one of the MDL algorithms. 
The SSC objective function~\erf{eqn:objective} combines the concerns of all three of these complexity measures. (This relation of SSC and complexity measures is returned to in Sec.~\ref{sec:info_flow_and_complexity} below.)

\subsection{Heuristics for minimizing the SSC objective function}
\label{sec:heuristics}
Because truly optimizing the SSC objective function---or doing some variant of the original multi-objective optimization---will often be quite difficult (if not formally uncomputable), there are several heuristics one might employ that could still yield advantages over the \latinit{ad hoc} nature of intuitive state space compressions. 

We already started on this path when we decided to focus on the situation where $\phi$ is first-order Markovian (that being a ``simpler'' dynamics to calculate than higher-order stochastic processes, of the type that are typically
used in time-series reconstruction using delay embeddings). 
An obvious next step---common in real-world instances of SSC, like those discussed in
Sec.~\ref{sec:related_work}---is to fix $Y$ ahead of time to some space substantially smaller than $X$, rather than try to optimize it. (In the case of Euclidean $X$, $Y$ will also be a Euclidean space of much smaller dimension; for finite $X$, $Y$ is also finite, but far smaller.) We may still optimize over $\pi$, $\phi$, and $\rho$, but in this heuristic the choice of $Y$ is fixed, decreasing the size of the search space dramatically.

Another heuristic that will often also make sense is to restrict the set of compression maps $\pi$ that are considered, for example, to some parametrized class of maps. In particular, when $Y$ is Euclidean,
we can restrict $\pi$ so that it cannot encode an arbitrary dimensional space $x \in X$ in an arbitrarily small dimensional $y \in Y$ with perfect accuracy, for example by restricting $\pi$ to a class of maps that are all continuously differentiable of a certain order, or Lipschitz. Without such restrictions, there will often be ``valid'' $\pi$ that depend sensitively on all infinitely many digits of $x$, such as the position along a space-filling curve; as a practical matter, such $\pi$ are impossible to compute. Even if we wish to consider $\pi$ that are distributions instead of functions, a parametrized family of distributions (e.\,g., a parametrized class of Gaussian processes) might prove fruitful. Optimizing the SSC objective function then amounts to optimizing over the parameter space of the chosen class of maps.


Note that to evaluate the minimum computation cost for a given map from $x_0$ to values $y_t$
would be equivalent to solving for the optimal
compilation of a given computer program down to machine code. In real computers, design of optimal compilers is still a very active area of research; calculating the cost
of such an optimized compilation will not be possible in general.{\footnote{Indeed,
even if we allowed an infinitely expandable RAM, such a cost would be uncomputable,
in general.}} Even calculating such costs for the abstracted version of real-world computers 
will likely prove intractable. However it should be possible to put bounds on such costs.
Moreover, purely pragmatically, one can run a search algorithm over the space of $\phi$'s, 
finding a good (if not optimal) compilation, and evaluate its associated cost.

Examples of heuristic ways to approach SSC are already present in related work, as discussed in Sec.~\ref{sec:related_work}.

\section{Applications of SSC -- beyond improving prediction} 
\label{sec:info_flow_and_complexity}

Although our SSC framework is most directly concerned with simplifying prediction of the future of a dynamical system, it may 
have other advantage for analyzing dynamic systems as well. 
As a particularly
simple example,  SSC provides a generic method of comparing systems: If two systems of the same type, e.\,g. two networks, or two economies have very similar optimal SSC compressions, this means that the ``underlying" dynamics of them
is quite similar. Intriguingly, this is true even for completely different types of systems: If an economy has some good state-space compression that very similar to a good state-space compression of some multicellular organism, that tells us that in
an important sense, that economy ``is" a multicellular organism, and vice-versa.

In the remainder of this section we present two other possible uses of SSC
beyond the domain of simplifying prediction that are somewhat more speculative.

\subsection{A form of complexity motivated by SSC}

Suppose we have a  compression of a system that
substantially reduces the value of the objective function $K$ compared to its  ``null
compression" value (i.e., the value for the identity compression that 
actually leaves the original space $X$ unchanged). Formally, 
using ``$\text{id}$" to indicate an identity map (with the spaces it operates varying and implicit), such a 
choice of $Y, \pi, \phi$, and $\rho$ results in a value of
\ba
\frac{ {K}(\pi, \phi, \rho; P) - {K}(\text{id}, \text{id}, \text{id}; P) }
   {{K}(\text{id}, \text{id}, \text{id}; P)} \; , 
\ea
close to $-1$ (assuming the objective function is guaranteed non-negative).

When this ratio is close to $-1$, the compression provides an emulator of the
dynamical system that is both easy to evolve computationally and
that accurately predicts the future of that dynamical system. When it is large however (i.\,e., only barely below zero), either the  emulation 
is not very compressed (i.e., computationally difficult to evolve) or it is
a poor predictor of the future of the underlying system, or both.
Accordingly, we define  \textbf{compression complexity} as
\ba
\frac{ \min_{\pi, \phi, \rho} {K}(\pi, \phi, \rho; P)}
   {{K}(\text{id}, \text{id}, \text{id}; P)}
\ea
ranging from 0 for highly compressible (non-complex) systems to 1 for highly incompressible (complex) systems. 
Note that compression complexity is defined \emph{with respect to a given accuracy cost
and associated observation operator}. So the same fine-grained system may be characterized as
``complex" or ``not complex" depending on what one wants to predict concerning its future
state. 

Compression complexity of a dynamical system is loosely similar to algorithmic information complexity (AIC)
of a string, viewed as a dynamical system. (Perhaps more precisely, it is similar
to techniques related to AIC that  allow ``noisy" reproduction of the string, e.g.,
minimal description length techniques.) However there are some important differences. To illustrate these,
recall that algorithmic information complexity (AIC) is high both for random strings and for ``complex'' strings.
In contrast, SSC compression complexity is low for random strings.
More precisely, consider the case where $\Omega = X$, $\mathscr{O}$ is the identify map,
and the fine-grained dynamics is IID  noise, i.e., $p(x_{t+1} \mid x_t) = p(x_{t+1})$ with a high entropy $H(X_t)$.
Suppose as well that accuracy cost is the time-averaged form of mutual information defined in~\erf{eq:mut_info_acc},
$-I_{\overline{\mathscr{P}}_{\pi,\phi,\rho}}(X' ; X)$.
For this situation, no choice  of $\pi, \phi$, and $\rho$ results in smaller
accuracy cost than when no compression is used. So if our SSC objective function were accuracy cost
alone, we would conclude that compression complexity is high (just like with AIC).

However consider compressing $X$ to a $Y$ that only contains a single element. We lose nothing in accuracy cost. But computation 
cost is now zero. So the value of the \emph{full} SSC objective function is
greatly reduced. This illustrates that ``a fine-grained dynamics that is IID 
noise" is assigned a small compression complexity.


When compression complexity is low, the compressed system $\{Y, \pi, \phi, \rho\}$ may provide substantial
physical insight into the dynamics $P$ of the microstate $X$, since the compressed system intuitively tell us ``what's important about $X$'s time-evolution''.
As an example, similar to Ex.~\ref{ex:2} above,
consider a never-ending (multi-stage) noncooperative game among many
players that models an economy where the players are employees of many firms interacting with one another.
Also posit a particular learning rule whereby the players jointly determine their dynamic behavior.
The dynamics of such a game is a stochastic process in a very high-dimensional space. However if its
compression complexity is low, then the 
optimal state compression of that dynamics provides a relatively small set of coarse-grained variables whose dynamics
captures the salient aspects of the full underlying game.{\footnote{Note that the optimizing $\pi$ in general
may be highly nonlinear, and that it may be that the optimizing $\phi$ cannot be interpreted as 
a multi-stage game involving the coarse-grained variables.}} Potentially each component of that optimal coarse-grained macrostate
is a characteristic of a separate one of the firms that are being modeled. Intriguingly though, it may well be that the macrostate
instead reflects characteristics of the joint behavior of multiple firms, 
or of sets of individuals spread across multiple firms. This would suggest that the proper way to
model this economy is not on a firm-by-firm basis, but rather in terms of macrostate variables that involve the states of multiple firms and individuals simultaneously. 

Analogously, this is precisely what happens in quantum mechanics: a description of a multi-particle system cannot merely describe the individual states of the particles, but must also include correlations between the particles, i.\,e. their entanglement.

As defined above, compression complexity is normalized, by the value ${K}(\text{id}, \text{id}, \text{id}; P)$.
This is not always appropriate, e.g., if there is some concrete physical meaning associated with the 
value of the objective function. In such cases, there are several possible modifications to the definition
of compression complexity that may make sense.
For example, in some situations it may be appropriate to quantify complexity
as $\min_{\pi, \phi, \rho} {K}(\pi, \phi, \rho; P)$. As always with SSC, the precise scenario
being analyzed should motivate the precise definitions that are used.

Unlike many other complexity measures, compression complexity is tailored to measuring
the complexity of \emph{dynamic} systems. Indeed, in its simplest form, it does not apply to static
objects like images. There are various modifications of compression complexity that can
apply to static objects though. For example, if we have a generative process that creates images,
we could measure the compression complexity of the generative process that produces that image.
%
%
%

We also note that in general, as a system evolves its optimal state space compression will change. 
So as we go from $t$ to $t+1$ to $t+2$, etc., if evaluate the associated
values of compression complexity taking each of those successive times as the ``initial time",
in general we would expect that complexity to undergo a dynamic process. This may provide a useful framework 
for analyzing informal suppositions of many fields concerning how complexity evolves
in time, e.g., concerning how the complexity of large biological
systems changes in time.

Finally, we emphasize again that we do not promote compression complexity
as ``the" way to measure complexity. Rather we are simply highlighting
that it has many aspects which match well to characteristics of complexity that have been
proposed in the past, and that it may lead to novel insight into physical phenomena.

\subsection{Using SSC to define information flow among scales of a system}

Many measures of ``information flow" in a stochastic dynamical system like
causal information flow~\cite{doi:10.1142/S0219525908001465}, transfer 
entropy~\cite{Schreiber:2000uq,prokopenko2014transfer,prokopenko2013thermodynamic}, and
Granger causality~\cite{kaminski2001evaluating} are motivated by considering the flow between \emph{physically distinct subsystems}
of an overall dynamic system. This makes them inherently ill-suited to quantifying
information flow among the scales of a {single} system, in which one would expect
there to be \latinit{a priori} ``biases" reflecting the fact that behavior at different scales of a 
single system cannot be completely independent. In fact, the complete opposite is typically
the case: the dynamics of the more detailed scale typically completely determine the dynamics of the less detailed scale. 

As an example of the
difficulties such biases cause for those measures of information flow, the most straightforward application
of transfer entropy (taking $k = l = 1$, in Schreiber's notation) would quantity the information flow from $Y_t$ to $X_{t+1}$ as the
conditional mutual information $I_{\pi, \phi ; P}(X_{t+1} ; Y_t \mid X_t)$. However this is identically zero, regardless of 
$\pi, \phi$, and $P$.  (Note that this is not exactly the case for the calculations of
transfer entropy between scales made  in~\cite{walker2012evolutionary}, since
they do not adopt this most straight-forward use of transfer entropy. However the same
kinds of statistical biases still apply.
From a certain perspective, these biases mean that the very notion of information flowing from a
high scale to a small scale of a single system is specious. 

Nonetheless, going back at least to the seminal work of Maynard-Smith and others,
researchers have presumed that information does ``flow" from the high
(coarse-grained) scale down to the low (fine-grained) scale of a biological system
organisms~\cite{Smith:2000fk,mayn70,walker2012evolutionary,davies2012epigenome}.
Indeed, Maynard-Smith averred that not only does such high-to-low scale information
flow exist, but that it has increased in each major transition in biological evolution. More recent work has also
emphasized the role that information flow between scales
may play in understanding the emergence of social complexity~\cite{dedeo_thisvol}
and even biological individuality~\cite{krakauer2014information} itself.

A similar phenomenon is seen in intuitive descriptions of computational algorithms, e.g.,
as implemented on Turing machines or with Von Neumann architectures. 
For example, consider an algorithm that determines whether its input is a prime number or not. On the one hand, the behavior of this algorithm is completely specified by its code: how it moves bits around in memory and combines them. On the other hand, the low-level bit-wise behavior of this algorithm may be viewed as being ``determined'' by its 
high-level characterization as a search for primality. When at some point the algorithm
takes one branch of a conditional ``if'' statement rather than another, do we say that it took that branch because the memory was just in a certain position, or because (to give an example) the number $6$ that was
input to the algorithm is not prime?

The SSC framework provides ways to formalize the information flowing from
a high scale down to a low  scale of a single system which are more directly grounded
in the relation between the behaviors at different scales of a single system than are measures
like transfer entropy.
The basic idea is to first use SSC to solve for the scale(s) at which to analyze the system, rather than rely on the scale(s)
being pre-specified in an ad hoc manner. We then define the associated information flow from the coarse scale to the fine
scale by treating ($\phi, \rho)$ as specifying an information channel between $Y$ and the observable function of $X$.
We refer to any such quantification based on
SSC as \textbf{inter-scale causality}, in analogy with ``Granger causality".

As an example of inter-scale causality, one might argue that the amount of information needed to construct 
the current value of $x$, given any particular estimate of it based on 
the value of $y$, 
is an appropriate measure of the ``information flow" from the coarse-scale to the fine
scale. This can be quantified as the entropy of $x_t$, evaluated
under the probability distribution conditioned on the value $y_{t-1}$. Expressing this
in terms of observables and predictions for them rather than directly in terms of $x$ and $y$, averaging over $y_{t-1}$,
and then averaging over $t$, we arrive at the measure
\ba
-\int d\Delta t \; W(\Delta t) H_{{\mathscr{P}}_{\pi, \phi, \rho ; P}}(\Omega_{\Delta t}\mid \Omega'_{\Delta t - 1}) 
\label{eq:info_flow_cond_entr}
\ea
(using the notation of Sec.~\ref{sec:info_theory_obj}).

A slight variant of this measure arises if we wish to normalize each of the conditional entropies
$H_{{\mathscr{P}}_{\pi, \phi, \rho ; P}}(\Omega_{\Delta t}\mid \Omega'_{\Delta t - 1})$ by subtracting it from $H_{{\mathscr{P}}_{\pi, \phi, \rho ; P}}(\Omega_{\Delta t})$. This difference tells us how much
extra information about the value of $\omega_{t}$ is provided by $\omega'_{t-1}$, beyond the prior information
concerning $\omega_t$. With this normalization our measure of information flow becomes
\ba
-\int d\Delta t \; W(\Delta t) I_{{\mathscr{P}}_{\pi, \phi, \rho ; P}}(\Omega'_{\Delta t}\; ; \Omega_{\Delta t}) 
\label{eq:info_flow_mutual_info}
\ea

This is just the time-averaged mutual information, a natural candidate for an information-theoretic accuracy cost. However
as discussed above, to avoid the problem that two variables can have perfect mutual information by being either perfectly correlated or perfectly anti-correlated at any given time step---and whether they are correlated or anti-correlated at each time step can change without changing the preceding value---it may make more sense to replace the above quantity with the mutual information of the time average, or even with the conditional entropy of the time average, $H_{\overline{\mathscr{P}}_{\pi,\phi,\rho}}(\Omega \mid \Omega')$. Any such
alternatives based on the concerns raised in the discussion of accuracy cost Sec.~\ref{sec:info_theory_obj} should be treated carefully though.
Here we are interested in quantifying information flow from the time series over $\Omega'$ (predicted from the time series over $Y$) to the time series over $\Omega$ (the observable applied to the time series over $X$). We are not interested in formalizing how accurately we can predict $\omega_t$ from $\omega'_t$ (or $y_t$).


It is worth emphasizing the essential difference between this kind of SSC-based measure of information flow between scales
and measures of it like the transfer entropy~\cite{Schreiber:2000uq} between  
scales. Measures like transfer entropy only depend on how much the time series $y_{t}$ tells 
us about $x_{t}$ that does not arise through direct dependence of $x_{t}$ on
$x_{t-1}$. For example, if $x_{t} \rightarrow y_{t}$ is a single-valued invertible mapping, then
$y_{t+1}$ will have high mutual information with $x_{t+1}$, even though there is no \emph{novel} information
that flows from $y_{t+1}$ to $x_{t+1}$, only different echoes of $x_{t}$. Transfer entropy is
explicitly designed to remove such effects arising from the direct dependence of $x_{t}$ on
$x_{t-1}$.

However this ``direct dependence" that transfer entropy is designed to ignore is  \emph{precisely} what  
we want to capture when considering information flow among scales. 
In particular, it is precisely what the measures suggested in~\erf{eq:info_flow_cond_entr} 
and~\erf{eq:info_flow_mutual_info} are designed to capture, due to how
${\mathscr{P}}_{\pi, \phi, \rho ; P}$ is defined.
Simply put, information flow between scales of a single system is fundamentally 
different from information flow between physically separate systems, and therefore should be quantified differently.

\section{Conclusions} \label{sec:discussion}

This preliminary report has presented a new framework for understanding how we construct higher-level descriptions of complex systems. We have introduced the problem through a series of illustrations, defined the key quantities, and provided three explicit examples of how this framework can be applied to basic problems in biology, economics, and physics. Having built an intuition for the method, we then compared this framework to a number of influential suggestions in the literature.

Our framework makes explicit two core questions for both the scientist and engineer: how accurate a theory is, and how difficult it is to work with. We have presented new theoretical results for how to quantify the answers to these questions, and how to combine them into a single objective function.

By formalizing the goals of a scientist engaged in providing a coarse-grained description of a system, state space compression allows us to compare and contrast a wide variety of systems. It provides novel ways to address long-standing problems that arise both within fields and between disciplines, where the question of ``how much to ignore'' becomes critical.

\section*{Acknowledgements} We thank Daniel Polani, Eckehard Olbrich, Nils
Bertschinger, Nihat Ay, Cris Moore, and James O'Dwyer for helpful discussion. We also
thank the Santa Fe Institute for support. In addition
this paper was made possible through the support of Grant No. TWCF0079/AB47 from 
the Templeton World Charity Foundation. The opinions expressed in this paper are those of the author(s) and do not necessarily reflect the view of Templeton World Charity Foundation. S. D. thanks the City University of New York's Initiative for the Theoretical Sciences for their hospitality during the course of this work. S. D.  was supported in part by National Science Foundation Grant EF-1137929. J. A. G. and E. L. acknowledge the support of Santa Fe Institute Omidyar Fellowships.

\providecommand{\bysame}{\leavevmode\hbox to3em{\hrulefill}\thinspace}
\providecommand{\MR}{\relax\ifhmode\unskip\space\fi MR }
\providecommand{\MRhref}[2]{%
  \href{http://www.ams.org/mathscinet-getitem?mr=#1}{#2}
}
\providecommand{\href}[2]{#2}

\end{document}